\documentclass[pdflatex,sn-mathphys,12pt,square,numbers,sort&compress,super]{sn-jnl}

\usepackage{times}

\usepackage{geometry}
\jyear{}%

\theoremstyle{thmstyleone}%
\theoremstyle{thmstyletwo}%

\theoremstyle{thmstylethree}%

\begin{document}

\title[Deformation-induced Topological Transitions in Mechanical Metamaterials]{Deformation-induced Topological Transitions in Mechanical Metamaterials and their Application to Tunable Non-linear Stiffening}

\author*[1]{\fnm{Marius} \sur{Wagner}}\email{marius.wagner@mat.ethz.ch}
\equalcont{These authors contributed equally to this work.}
\author*[1]{\fnm{Fabian} \sur{Schwarz}}\email{fabian.schwarz@mat.ethz.ch}
\equalcont{These authors contributed equally to this work.}

\author[1]{\fnm{Nick} \sur{Huber}}

\author[1]{\fnm{Lena} \sur{Geistlich}}

\author[1]{\fnm{Henning} \sur{Galinski}}

\author[1]{\fnm{Ralph} \sur{Spolenak}}

\affil*[1]{\orgdiv{Laboratory for Nanometallurgy, Department of Materials}, \orgname{ETH Zürich}, \orgaddress{\postcode{Vladimir-Prelog-Weg 5, CH-8093}, \city{Zürich}, \country{Switzerland}}}

\abstract{Mechanical metamaterials are periodic lattice structures with complex unit cell architectures that can achieve extraordinary mechanical properties beyond the capability of bulk materials. A new class of metamaterials is proposed, whose mechanical properties rely on deformation-induced transitions in nodal-topology by formation of internal self-contact. The universal nature of the principle presented, is demonstrated for tension, compression, shear and torsion. In particular, it is shown that by frustration of soft deformation modes, large highly non-linear stiffening effects can be generated. Tunable non-linear elasticity can be exploited to design materials mimicking the complex mechanical response of biological tissue.
}

\keywords{Mechanical Metamaterials, Topology Transition, Kinematically Indeterminate Frameworks, Non-linear Stiffness, Strain-Stiffening, Deformation Modes, 3D Printing}

\maketitle

\section{Introduction}\label{sec1}

\begin{figure}[t!]
    \centering
    \includegraphics[scale=1]{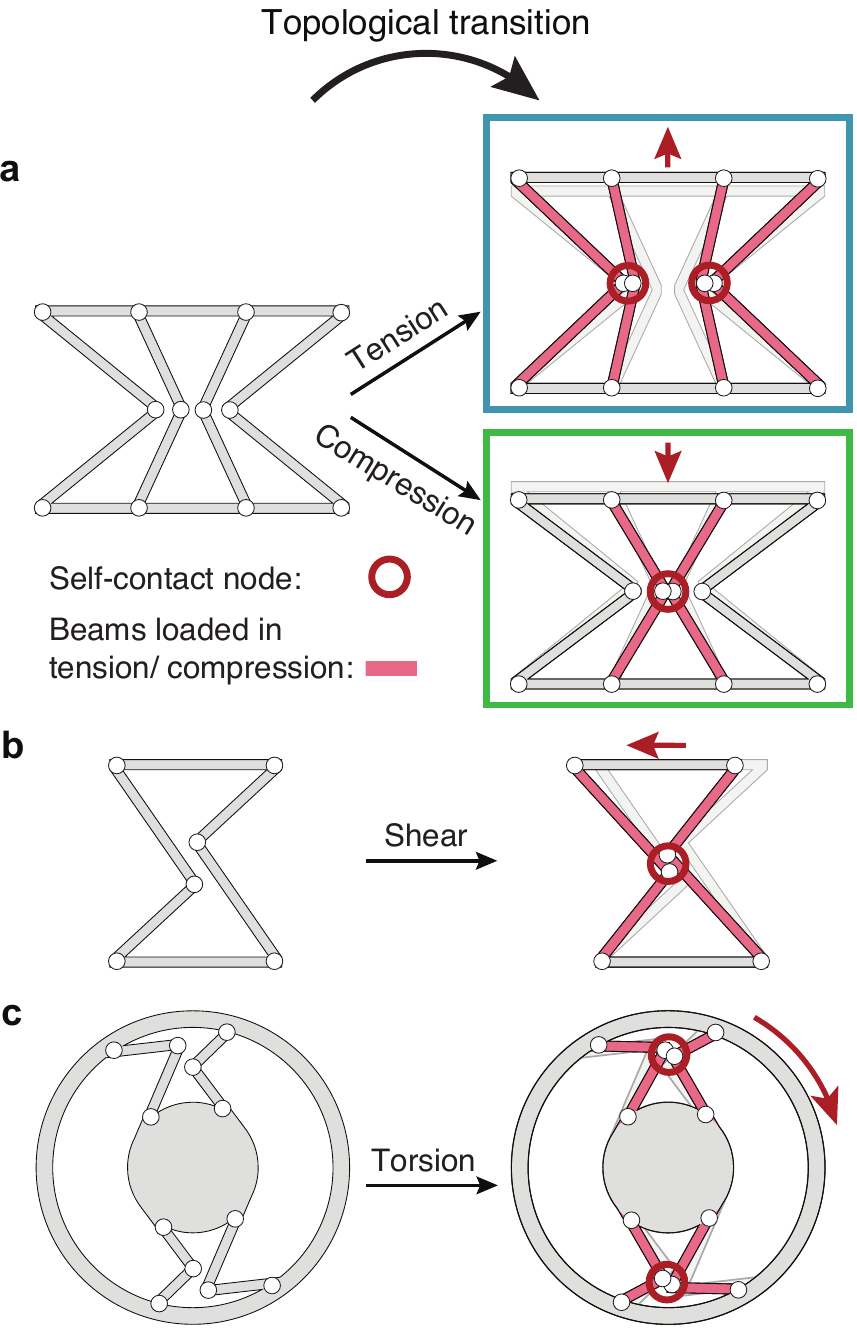}
    \caption{\textbf{Schematic unit cells of the metamaterials, left column initial undeformed state, right column reconfiguration of nodal-topology.} When deformed in tension and compression \textbf{(a)}, shear \textbf{(b)}, or torsion \textbf{(c)}, internal contacts form. The two contacting nodes transmit loads and hence, effectively behave like one single node (red circles). The transition in connectivity leads to a change in the loading conditions of the beams from bending to tension and compression (highlighted red). The red arrows are indicating deformation.}
    \label{fig:1}
\end{figure}

Biological tissues, like skin \cite{Rn45}, tendons \cite{Rn48}, or inter-vertebral disks \cite{Rn49,Rn57} exhibit dynamic non-linear elastic properties. While at small external forces these biological materials are soft and deformable, they strain-stiffen at higher external loads. This mechanical characteristic is fundamental for the physiological function by preventing damage from excessive strain \cite{Rn43,Rn62}. Although the design principles of these biological systems are far from being fully understood, the physical origin of strain-stiffening has been successfully attributed to the connectivity of the cytoskeletal network and the straightening of chains, which constitute the mechanical backbone of these tissues \cite{Rn65,Rn63,Rn61}.\\
Owing to the potential for regenerative medicine, there has been substantial effort in synthetic semi-flexible polymer networks \cite{Rn66,Rn64,Rn63,Rn58,Rn59} and mechanical metamaterials \cite{RN1_Ex,RN2_Ex,RN3_Ex,RN4_Ex,Rn60} in which the remarkable characteristic of biological tissue may be duplicated. 
Thereby, mechanical metamaterials offer an unique design space to control the mechanics and connectivity at the unit cell level.\\
So far, two fundamental approaches have been used to realize mechanical metamaterials with non-linear elastic properties. The first employs wavy networks of soft polymer, which straighten upon stretching \cite{Rn60}. As members become straight, their internal loading condition switches from bending to tension, resulting in stiffening. The second approach~\cite{RN1_Ex,RN2_Ex,RN3_Ex,RN4_Ex} uses the intuitive idea to form internal self-contacts under compression. The newly formed contacts change the path of load-transmittance. Elements which become stressed will reduce the strain from additional external loads and hence increase the material's resistance to deformation.\\ 
However, non-linear elasticity in mechanical metamaterials is currently limited to a single loading condition i.e. tensile loading for soft network metamaterials and compression for contact-forming metamaterials. This emphasizes the lack of a generalized principle to universally generate non-linear elasticity.\\
\\
This work presents mechanical metamaterials which overcome these limitations, introducing a generalized approach to generate contact-induced strain-stiffening independent of the loading condition.
We start from the observation that the connectivity of cellular solids, i.e. the nodal-topology, presents an essential factor for its resistance to deformation \cite{RN25}. Consequently, changes in the nodal-topology, as they are found in the class of contact-forming metamaterials mentioned above, will result in non-linear elasticity. While those studies take advantage of this principle, the mechanical response observed empirically, is not discussed in the context of nodal-topology.
\\
So far, the use of topology as design parameter, has been mostly limited to metamaterials that exhibit a single topological configuration, which is invariant under deformation. Only a few state-of-the-art systems~\cite{RN22,RN16,RN12} attempt to expand the space of accessible properties by introducing topological transitions, but these systems are either bound to a substrate~\cite{RN16} or the topological transition upon deformation remains limited to a single loading condition~\cite{RN22,RN12}.\\ 
\\
In this study, we draw inspiration from stiffness-tuning by contact-formation and significantly expand the capabilities of the state-of-the-art by achieving changes in nodal-topology under arbitrary loading conditions. This is demonstrated for tension, compression, shear and torsion. Further, the fundamental mechanism of the stiffening, the change in nodal-topology, i.e. the connectivity between the real-space elements, is examined. To maximize the effect of the connectivity reconfiguration on the elastic non-linearity, the metamaterials are designed to switch from a soft - kinematically indeterminate initial state to a rigid - determinate state when deformed.\\
Soft inextensional modes of deformation in kinematically indeterminate frameworks allow for large reconfigurations by rotation or bending in joints without inducing tension or compressing in the beams.
These soft modes are employed to design the displacement trajectory of nodes connecting beams.
Upon deformation inter-nodal contact is formed. For the given loading condition, the nodes in contact mutually constrain the displacement of their counterpart, which causes frustration of the soft deformation mode. Further deformation requires changes in the beam length (rigid, stretch-dominated deformation) instead of rotations in
joints (soft, bending dominated deformation).
Consequently, the contact-pairs, can be regarded as one single node for the given loading conditions and the effective connectivity, i.e. the nodal-topology of the framework is altered, resulting in a drastic, discontinuous increase in elastic modulus.\\
The proposed concept, illustrated in Fig.~\ref{fig:1}, does not require any external trigger, but exclusively harnesses the design of the metamaterial. Furthermore, it can be applied to universal loading conditions. As the transition takes place in the elastic regime of the constitutive material, complete reversibility is achieved.\\
It needs to be stressed that the phenomenon leading to the change in modulus is vastly different from the compaction of cellular solids \cite{RN25}. The contact formation in this work takes place at precisely controlled external loading conditions, is not limited to compression, and is fully elastic.
\\
The stress and strain at which the transition in connectivity takes place, as well as the magnitude of the modulus change, can be controlled by rational design of the unit cell. More complex mechanical behaviors can be obtained by combining unit cells with different geometries. Numerical simulations are used to investigate the change in effective modulus and extensively explore the design space of the tension and compression type metamaterial (Fig. \ref{fig:1}a). The presented concept is validated experimentally by mechanical testing of selected 3D-printed structures in tension and compression.

\section{Methods}\label{sec4}
\subsection{Numerical simulations}\label{subsec4,1}
Numerical simulations of the mechanical metamaterials were performed using the solid mechanics module in COMSOL Multiphysics\textsuperscript{\tiny\textregistered} v. 5.6 \cite{comsol}. The contact was modeled using the penalty method. The systems were modelled using a 2D model, except for the systems that were compared with the experiment, which were simulated in 3D. Due to the 2D nature of our systems, the 2D FEM model is sufficient to simulate them. The mesh was chosen after a mesh sensitivity analysis. In 2D, triangular elements were used, in 3D tetrahedral elements were used accordingly. The individual elements were of quadratic serendipity discretization. The average system of 3x3 unit cells had about $1-5 \times10^5$ degrees of freedom and took about $20-30$ minutes to compute using 4 processors with a clock speed between 2.3 and 3.7 GHz on the ETH Zürich Euler cluster (\url{https://scicomp.ethz.ch}).\\
\\
A hyper-elastic Mooney-Rivlin material model with 5 parameters was used. The parameters were fitted to experimental data. The stiffness was obtained by statically applying a step-wise displacement at the top boundary and measuring the reaction force. A roller boundary condition was applied to the bottom edge, allowing movement in x-direction, and constraining displacements in y-direction, in which deformation is applied. Rigid body translations in x-direction were prevented by fixing the left bottom corner. The effective nominal stress and strain were calculated by dividing the applied displacement by the initial height, respectively the reaction force by the crossectional area in the initial state or in the state at the point of contact formation. The multi-frontal direct sparse MUMPS solver was used in all numerical calculations.
\subsection{Experimental characterization}\label{subsec4,2}
Specimens for experimental characterization were fabricated by Spectroplast AG, Switzerland using stereolithography (SLA) 3D printing of silicone (TrueSil A60). In this study the base material was chosen to be a silicone polymer. The advantage of silicone is its large purely elastic stains and small hysteresis and creep. The SLA 3D printing allows the fabrication of fine features with high resolution and good tolerances. Generally, our metamaterials can be fabricated from various materials. The condition for a reversible change in topology is that the maximum local strain remains in the elastic regime of the constitutive material. According to the numerical simulations, the maximum local strain at contact is in the same order than the effective strain applied, i.e. between 0.1\% and 10\%. Hence, also metals and polymers with a reduced elastic regime can be used. At the small scale that may also apply to semiconductors and ceramics.\\
\\
Compression and tensile tests were performed on an Instron ElectroPuls E3000 universal testing machine equipped with a 250 N Dynacell load cell, using a strain rate of 5\% min$^{-1}$. 3D-printed adapters (fused filament fabrication of PA6) were glued to the top and bottom surface of the specimen to enable gripping. The exact geometry of the tested specimens can be found in the Supplementary Information.

\section{Results}\label{sec2}

\subsection{Soft to rigid transitions in framework metamaterials}\label{subsec2,1}

\begin{figure}[t!]
    \centering
    \includegraphics[scale=1]{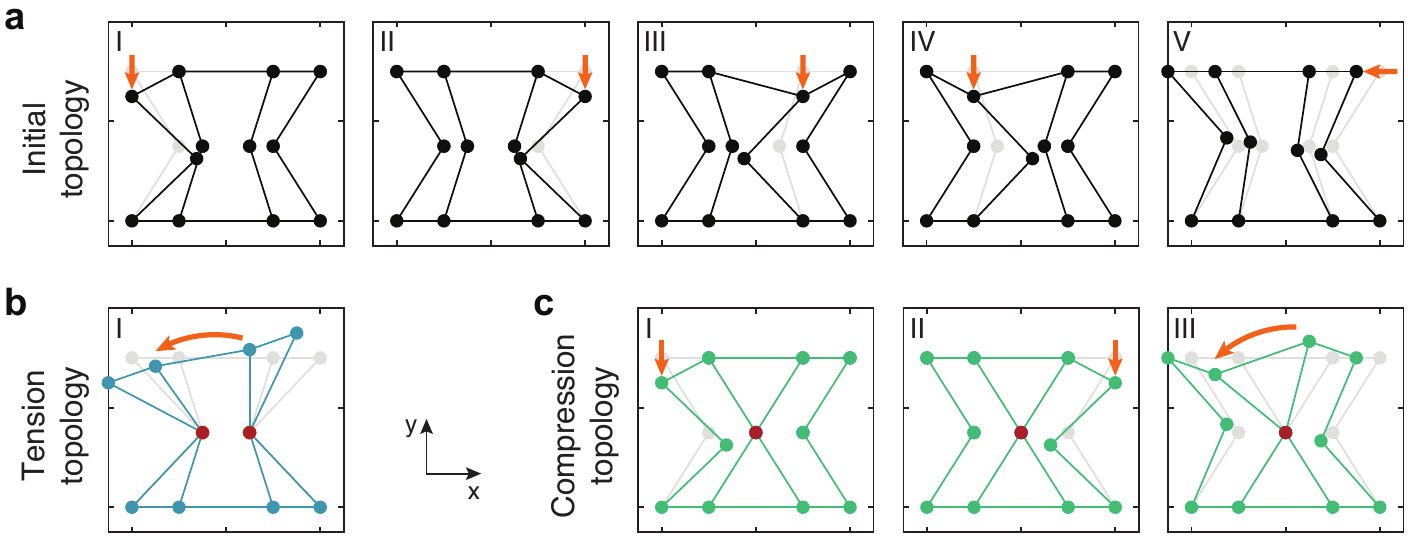}
    \caption{\textbf{Computed soft modes of the different framework topologies.} The contact-pairs formed, changing the nodal connectivity, are indicated in red. \textbf{(a)} The undeformed framework exhibits five modes of inextensional deformation. Straining in the y-direction excites mechanism I – IV, and the beams are not subjected to any loading. \textbf{(b)} In the tensile state, one  soft mode exists, which is not exited when tensile load in the y-direction is applied. This results in a rigid deformation mode. \textbf{(c)} The framework arising from compression exhibits three inextensional modes. When the framework is compressed in the y-direction, the inner beams are loaded in compression, leading to a rigid deformation behavior.}
    \label{fig:2}
\end{figure}
To predict the switch from a soft, to a rigid deformation mode in our metamaterials, we employ an extension of Maxwell's mechanical stability theory of frameworks \cite{RN36} developed by Pellegrino \textit{et al.} \cite{RN26}. For this analysis, the structure is approximated as framework of rigid beams connected by friction-less joints.
The soft modes of deformation and the states of self-stress are computed for a single unit cell in tension and compression (Fig. \ref{fig:1}a). The three connectivity-states of the framework (before contact, tension, and compression) are analyzed separately. Details can be found in the Supplementary Information. Constraints are applied to the lower edge of the structure to prevent any translations and to model the behavior of the structure being clamped on the lower edge. It is to note that the structure of interest does not contain any states of self-stress in the three states.\\
Fig. \ref{fig:2} illustrates the identified soft modes in the respective states. The graphs show the coordinates of the joints displaced by the force components, which excite soft modes. In the undeformed state, the framework possesses five soft modes (Fig. \ref{fig:2}a). When deformation along the y-axis is applied, mode I to IV are active, hence, no loading is induced in any beams. Mode V is associated with shearing of the unit cell. In the tensile state (Fig. \ref{fig:2}b), two self-contact nodes are formed and only one soft mode remains. When loaded in tension, this remaining mode is not active. Hence, the structure behaves rigidly, and the beams are subjected to tensile and compressive loading. When compressed along the y-axis (Fig. \ref{fig:2}c), the structure exhibits one self-contact joint and three soft modes. Mode I and II are active, however, the motion of the two top-middle joints, requires compression of the inner beams, resulting in an overall rigid deformation mode.\\

\subsection{Stiffening induced by transitions in nodal-topology}\label{subsec2,2}

\begin{figure}[t!]
    \centering
    \includegraphics[width=1\textwidth]{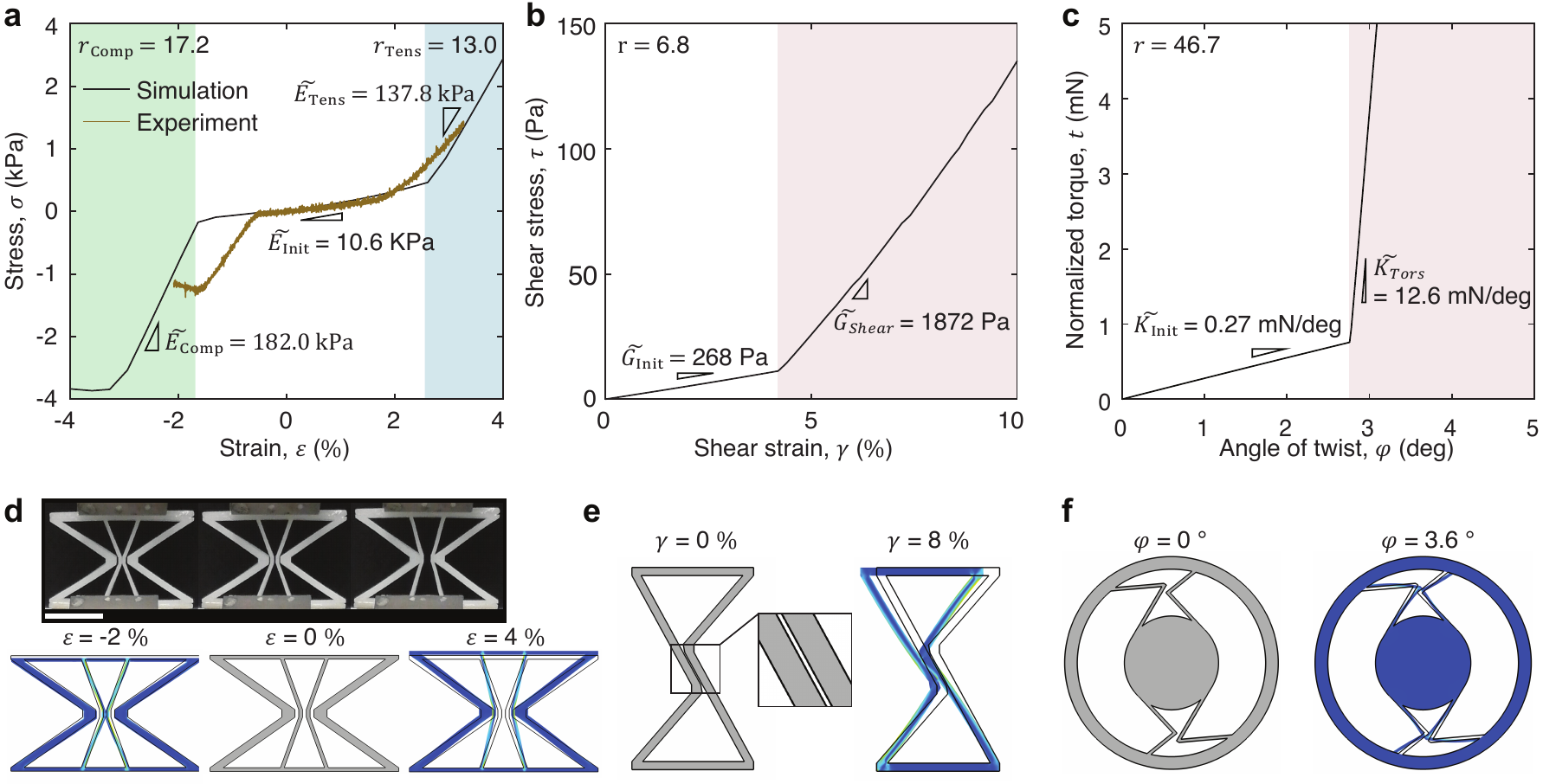}
    \caption{\textbf{Mechanical response of the connectivity changing metamaterial unit cells.} The change in nodal-topology manifests as discontinuity in the slope, i.e. the modulus \textbf{(a)}, \textbf{(b)} respectively stiffness \textbf{(c)}. The soft deformation mode of the initial kinematically indeterminate structure results in a low modulus. When the connectivity changes the deformation mode switches and a drastic increase in the slope of the stress-strain curve can be observed. The ratio between the moduli before and after contact is given as $r$. The numerical simulations show the states corresponding to the regimes of the stress strain curves in tension and compression \textbf{(d)}, shear \textbf{(e)} and torsion \textbf{(f)}. The colors are indicating the von Mises stress. A larger image of the simulated stress states can be found in the Supplementary Information. The 3D-printed specimen shows the same states of nodal-topology than the simulations \textbf{(d)}. Scale bar, 2 cm. The experimental stress-strain curve in tension and compression shows good qualitative agreement with the simulation \textbf{(a)}.}
    \label{fig:3}
\end{figure}

We examine the mechanical response of the metamaterials undergoing connectivity transitions by finite element modelling. All three unit cells show a drastic increase in effective modulus (Fig.~\ref{fig:3}a, b), respectively stiffness (c), as a result of the formation of self-contacts. The simulated structures are shown in Fig.~\ref{fig:3}d-f. The color indicates the normalized von Mises stress, which was chosen to visualize the complex loading conditions in the beams.\\
\\
To experimentally validate our concept, we compare the mechanical response of a 3D-printed silicone specimen to the numerical simulations (Fig.~\ref{fig:3}a and d).
The structure changes its nodal-topology under uniaxial deformation. When tensile deformation is applied, the inner nodes displace outwards faster than the outer ones. When compressed, the two central nodes move inward until contact is formed (see Fig.~\ref{fig:3}d). Subsequently, the beams are subjected to compression instead of bending until buckling is observed. Both the experiment and numerical simulation show a discontinuous increase in modulus upon formation of the internal self-contacts (Fig.~\ref{fig:3}a), clearly indicating a change in the deformation mode. The simulated and experimentally measured moduli in the different states are in good agreement, while the strain of contact differs substantially. This can potentially be attributed to differences in the inter-nodal spacing (Fig.~\ref{fig:3}d) of the initial undeformed state.\\
\begin{figure}[t!]
    \centering
    \includegraphics[width=1\textwidth]{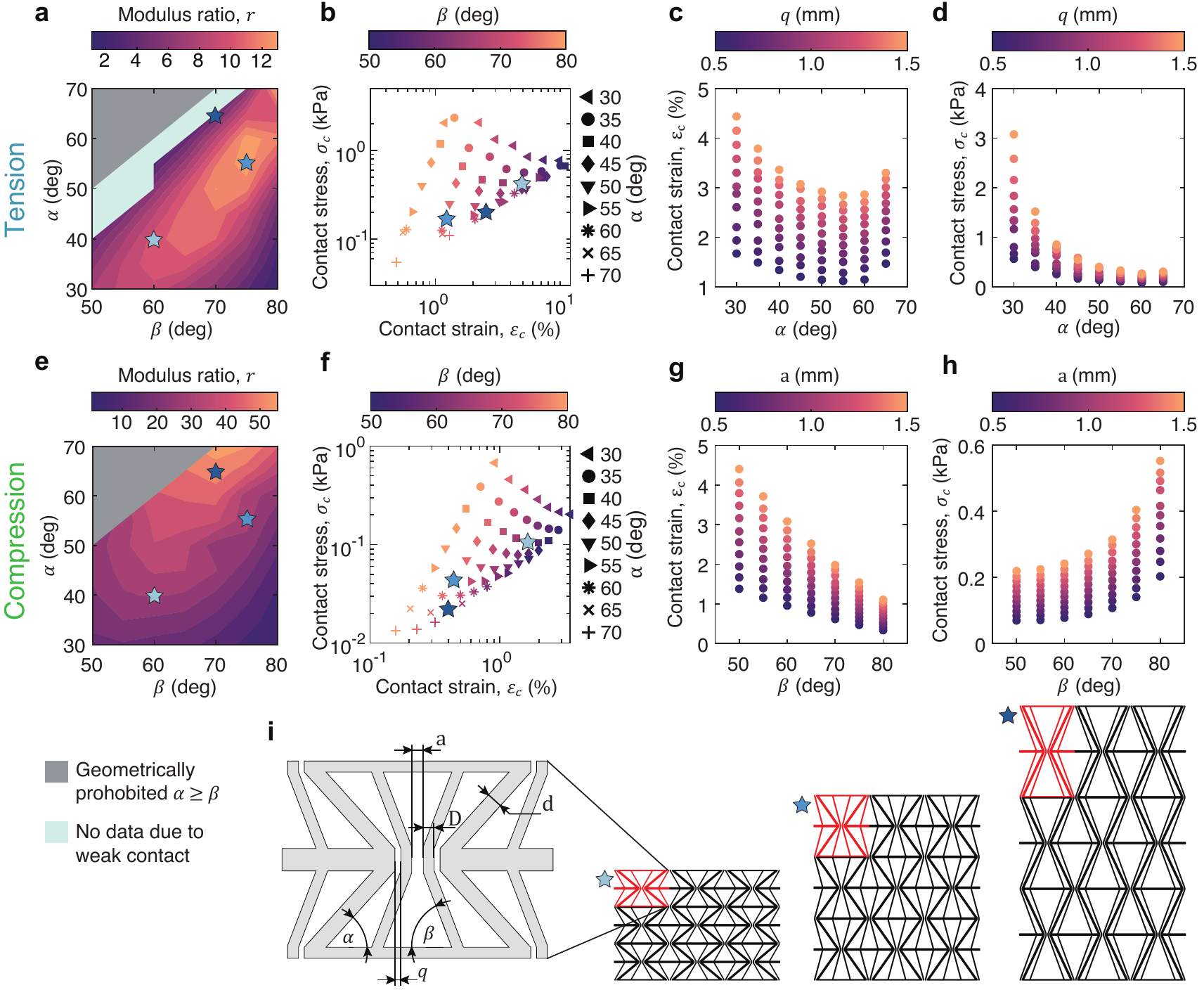}
    \caption{\textbf{Numerical exploration of the design space and the tunability of the topology-induced stiffening in tension (a - d) and compression (e - h).}  The modulus ratio can be controlled over a large extent by the angles $\alpha$ and $\beta$ \textbf{(a)}, \textbf{(e)}. Similarly, the stress and strain at which the topology transition occurs can be tuned by the two angles \textbf{(b)}, \textbf{(f)}, or by one angle and the inter-nodal distance $q$ \textbf{(c)}, \textbf{(d)} and  $a$ \textbf{(g)}, \textbf{(h)}. This allows to tailor the mechanical response over an impressive range of properties. \textbf{(i)} Single unit cell with the design variables analyzed in the study, and three of the simulated structures in the undeformed state (from left to right: $\alpha$ = 40°, $\beta$ = 60°; $\alpha$ = 55°, $\beta$ = 75°; $\alpha$ = 65°, $\beta$ = 70°).}
    \label{fig:4}
\end{figure}
Fig. \ref{fig:3}b and e report a metamaterial unit cell exhibiting a transition in connectivity when sheared. As the deformation is applied, the inner nodes approach until contact forms. Finally, Fig. \ref{fig:3}c and f illustrate a design forming inter-nodal contacts when torsional deformation is applied. Contact-pairs form when the outer ring is rotated clockwise, leading to a switch from a soft to a rigid deformation mode. All load cases presented show a substantial stiffening effect as a result of the transition in nodal-topology (Fig. \ref{fig:3}a-c). The stiffening can be characterized as ratio $r$ of the effective modulus before contact $\tilde E_{\text{Init}}$ and after contact $\tilde E_{\text{Trans}}$:
\begin{linenomath}\begin{equation}
r= \frac{\tilde E_{\text{Trans}}}{\tilde E_{\text{Init}}} \label{eq3}
\end{equation}\end{linenomath}
The ratio $r$ serves two purposes. First, it allows comparison of the stiffening in different metamaterials and design variable combinations. Second, it can be used as an indicator for a transition in deformation mode. If $r\gg1$, load-carrying internal contacts were established and the connectivity has effectively changed.\\
\\
The three types of unit cells presented, can be assembled into periodic metamaterials. In the following the metamaterial of the tension / compression unit cell will be further characterized. Tessellations of the shear and torsion unit cells into metamaterials can be found in the Supplementary Information.\\
The change in the nodal-topology can be described by means of the one-dimensional Betti number. Betti numbers are mathematical measures to distinguish topological phases \cite{RN41}. When applied to the planar representations of our metamaterials, the change in the nodal-topology can be described by the one-dimensional Betti number $b_1$, which equals the number of holes in the framework. 
As contacts are formed, the number of holes, i.e. $b_1$ increases.
We find that independent of the load case, an increase in the Betti number, given by positive $\Delta b_1$, relates to a change in the observed mode of deformation. A positive $\Delta b_1$ is a necessary but not sufficient condition for such changes. Further, $\Delta b_1$ can be used to compare the transition in nodal connectivity in different metamaterials. A detailed application of this concept to our metamaterials can be found in the Supplementary Information.\\
The unit cell of the tension / compression metamaterial investigated previously, is adjusted, such that the number of intra and inter unit cell contact-pairs and hence the effect on the mechanical response is maximized (Fig. \ref{fig:4}i). A comparison of the simulated mechanical response of a metamaterial consisting of the simple and the modified unit cell can be found in the Supplementary Information. This modification can be also motivated by the change in Betti number.
$\Delta b_1$ of the adjusted unit cell increases from 1 to 2 in compression and from 2 to 4 in tension. This also reflects in the magnitude of the nonlinearity of the elastic response, as it will be discussed in the following and it is highlighted in Supplementary Information Fig. S4.

\subsection{Design space of the mechanical response}\label{subsec2,3}

\begin{figure}[H]
    \centering
    \includegraphics[width=1\textwidth]{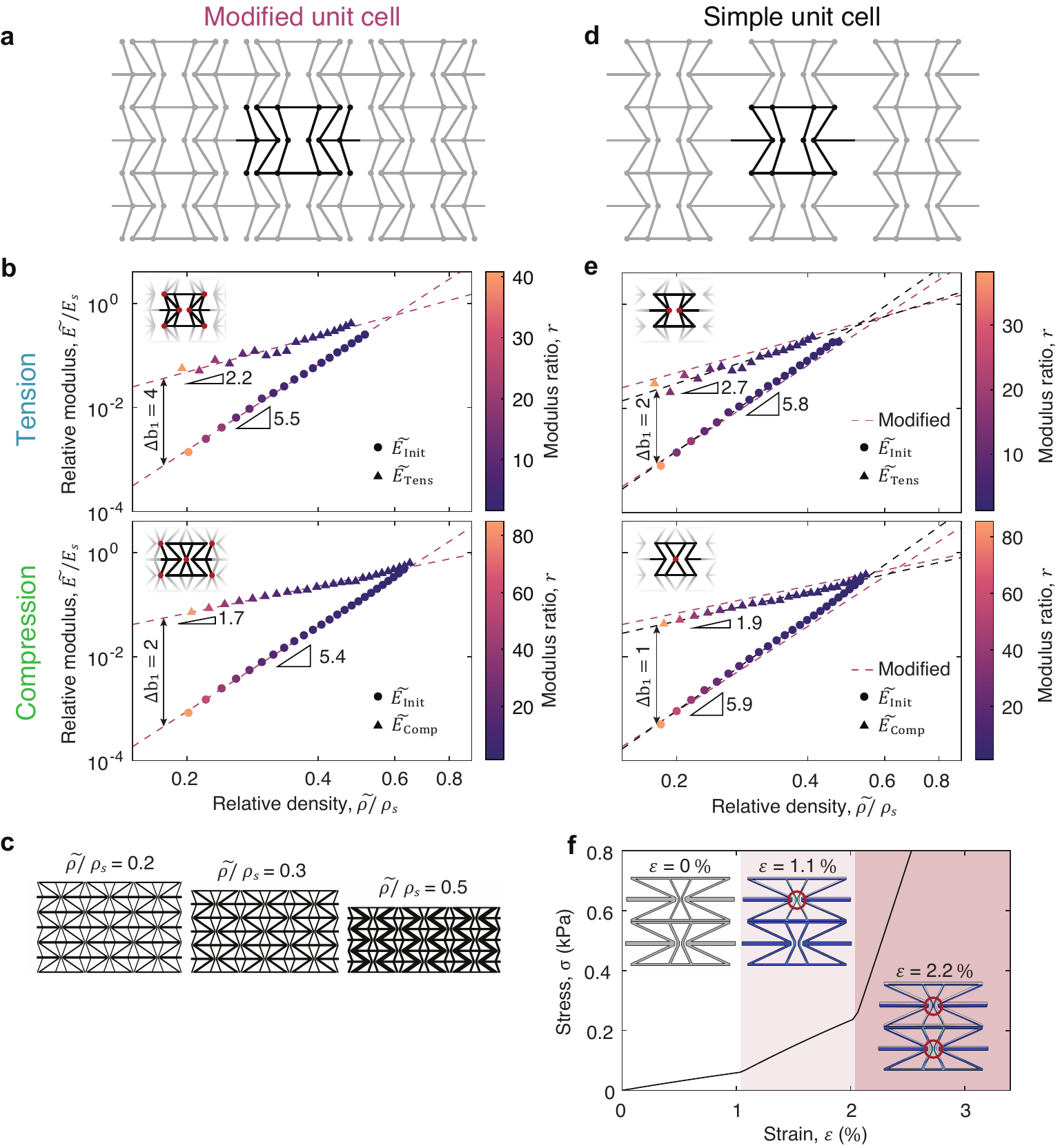}
    \caption{\textbf{Relative modulus of the tension and compression metamaterials scaling with the relative density, and exploration of a two-unit cell system exhibiting two sequential connectivity transitions.}
    \textbf{(a)} Illustration of the modified unit cell metamaterial, as used for the investigation of the design parameter influence. \textbf{(b)} Ashby graphs of the modified unit cell metamaterial in tension and compression. Different slopes indicate changes stress states due to topological reconfiguration of the structure. With increasing relative density, a decreasing modulus ratio $r$ is found. \textbf{(c)} Simulated structures of different relative density. \textbf{(d)} Schematic of a metamaterial using the simple unit cell. \textbf{(e)} Ashby plot of the simple unit cell metamaterial. A lower modulus ratio as well as larger slopes are found. \textbf{(f)} System of two-unit cells in serial configuration which sequentially change their topology. This grants further freedom in the design of the strain-stiffening behavior.}
    \label{fig:5}
\end{figure}

To determine the fundamental limits of our metamaterial, we explore the design space for the specific example of a metamaterial under tension and compression  (Fig. \ref{fig:3}a, d). 
\\
We use three different figures of merit to characterize the system, the modulus ratio $r$, the strain at contact $\varepsilon_c$ and the stress at contact $\sigma_c$. In Fig.~\ref{fig:4} we analyze the impact of the angles between the beams $\alpha$ and $\beta$ as well as the inter-nodal spacing $(a,q)$ on the mechanical response, while in Fig.~\ref{fig:5} the impact of the relative density $\rho$ is addressed. The parameterized unit cell with all design variables can be found in the Supplementary Information Fig. S6.
\\
The stiffening can be widely tuned by controlling the angles between the beams $\alpha$ and $\beta$ (Fig. \ref{fig:4}a and e). In the case of tension (Fig. \ref{fig:4}a) $r$ can be designed over 1 order of magnitude, while in the case of compression  (Fig. \ref{fig:4}e) 1.5 orders of magnitude are accessible.
Fig. \ref{fig:4}b and f show the change in the strain and stress at contact for the same combinations of $\alpha$ and $\beta$.
Additionally, $\varepsilon_c$ and $\sigma_c$ can be engineered by changing the inter-nodal distance $q$ and $a$ in combination with $\alpha$ or $\beta$ for both tensile (Fig. \ref{fig:4}c and d) and compressive loading (Fig. \ref{fig:4}g and h).\\
\\
Different cellular solids can be compared in a double logarithmic plot of the relative modulus against the relative density (Ashby graph), in which they typically show a linear scaling~\cite{Rn23}. The slope allows to draw conclusion of the stress state in the material~\cite{Rn55}. Here, the relative density, i.e. the volume of the constituting beams divided by the structure’s boundary, was varied as function of the beam thicknesses (design parameters $d$ and $D$, see Fig. \ref{fig:4}i), while the ratio between them was kept constant at 2:1 (Fig.~\ref{fig:5}c).\\
First, the metamaterial of the modified unit cell, which was also used for the exploration of the design space, will be discussed (Fig.~\ref{fig:5}a-c). We find a difference in scaling for the initial modulus (5.5 in tension, 5.4 in compression) and the transformed modulus (2.2 in tension, 1.7 in compression). As the slopes directly depend on the stress-state in the material, this difference highlights a fundamental change of the internal transmission of loads.\\
It can be shown that ideal stretch-dominated structures exhibit a slope of 1 and ideal bending-dominated structures scale with a slope of 2. Ideal in this context means a perfectly uniform stress in the entire structure. Heterogeneous stress distributions result in larger slopes~\cite{RN27}. This is also evident in our metamaterials, as they are architected to maximize the nonlinearity in elasticity and not to optimize the load carrying capability.\\
The change in the one-dimensional Betti number $\Delta$b$_1$ indicates the change in the connectivity and consequently the loading conditions leading to the change in slope between initial and deformed states.
\\
Note that the scaling will result in a relative modulus greater than one for a relative density of one. This stems from the normalization of the effective moduli with the initial tangent modulus. While a hyper-elastic progressive materials model is employed in the simulations.\\
In compression, $r$ can be tuned over almost two orders of magnitude, from r = 1.3 to r = 85 and in tension over almost one and a half orders of magnitude, from r = 1.6 to r = 41.
The minimum relative densities were selected such that a fabrication with the employed stereo lithography 3D printing process is possible ($D\geq$ 0.5 mm). The maximum relative densities at which a change and modulus was detected, and thus at which an effective transition in deformation mode is observed, is 0.5 in tension and 0.6 in compression.\\
The metamaterial with the simple unit cell (Fig.~\ref{fig:5}d) is compared to the metamaterial with the modified unit cell (a). We find a smaller difference between the initial modulus and the deformed modulus ($r$-ratio) both in tension and compression (Fig.~\ref{fig:5}e). 
 Fig.~\ref{fig:5}e shows the scaling for both unit cell types in tension (top) and compression (bottom). 
The moduli of the simple unit cell metamaterial generally scale with a larger slope. This indicates an overall more heterogeneous load distribution in the material. This is an intuitive observation, as the difference in stress in the diagonal beams compared and the horizontal ones is larger. Further, lower $r$ ratios are found. These findings highlight that the modification of the unit cell, associated with an increase in $\Delta$b$_1$, enlarged the non-linear elasticity of the metamaterial. The larger number of topological reconfigurations allow for more beams transmitting loads contributing to a rigid mode of deformation.\\
\\
More complex mechanical responses can be accessed by combining unit cells with different geometries into one metamaterial.
As it was shown in Fig. \ref{fig:4}, the contact stress and strain can be precisely adjusted over a broad range by selection of the design parameters. This enables the design of metamaterials, in which contact formation occurs at different stresses or strains, resulting in locally confined sequential topology transitions. Fig. \ref{fig:5}f shows a numerically simulated system of a two unit cell stack loaded in compression. The upper unit cell has a lower contact stress than the bottom one. Consequently, the upper unit cell switches from a soft to a rigid deformation mode first. Further deformation of the system leads to predominant straining of the lower unit cell to the point, at which it also changes the nodal-topology. The result is a stress-strain curve with two highly non-linear discontinuous changes in slope. Note that the simple unit cell is used in the stack configuration. The modified unit cell would not improve the topology-induced stiffening since no inter-unit cell contacts can be formed, as there are no laterally adjacent unit cells.\\
Additionally, the tunability of the contact strain can be harnessed to design meta-structures with programmable macroscopic deformation (Supplementary Information Fig. S7).

\section{Conclusion}\label{sec3}
In this work, we have introduced and characterized a new class of metamaterials, which rely on deformation-induced topological transitions in kinematically indeterminate frameworks, providing a new pathway to design materials with highly non-linear mechanical properties.\\
This general concept can be applied to arbitrary loading conditions, and a wide range of constituent materials.
While the presented unit cells are all created by rational design, a recent study introduced a mathematical approach to design infinitesimal or finite motion in kinematically indeterminate mechanical systems of joints connected by freely rotating rods \cite{Rn42}. This can be applied to design trajectories of nodal-displacement in mechanical metamaterials under arbitrary loading scenarios.\\
The change in deformation mode of our metamaterials can be predicted analytically, proving that the non-linear elasticity is associated with the frustration of soft modes.
Numerical simulations confirm the transition in nodal-topology and the resulting change in modulus for all load cases.\\
Experimental testing in tension and compression is in excellent agreement with the analytical and numerical models. Further, we propose that the change in the one-dimensional Betti number can be used as a metric for describing the change in the nodal-topology.\\
\\
The presented concept inherits the limitation that beams undergo small bending deformations, when deformed in the initial state. Consequently, the assumption of perfectly straight beams loaded in pure tension and compression, is not fulfilled completely. This is especially relevant if large nodal displacements are required to form internal contact. This issue can be mitigated by localizing the deformation using hinges \cite{RN28,RN34,RN35}.\\
\\
The stress and strain at which the topological transition takes place, as well as the magnitude of the stiffening can be tuned over almost two orders of magnitude by the unit cell design, as shown by a comprehensive numerical study of the design space.
The presented principle of universal topology-induced stiffening significantly extends the capabilities of the existing metamaterials \cite{RN2_Ex, RN3_Ex, RN1_Ex}.\\
\\
Our approach of generating highly tunable non-linear strain-stiffening can be used to design metamaterials emulating the non-linear elasticity of biological tissues~\cite{Rn43}. Examples include the non-linear response of human skin~\cite{Rn45} and tendons~\cite{Rn48} to tensile loading, the compressive and torsional stiffness of human intervertebral discs~\cite{Rn49}, or the strain-stiffening of soft biological tissue under simple shear deformation~\cite{Rn46}. Hence, the principle presented may find applications as bio-metamaterials~\cite{Rn44} in fields like tissue engineering or biomedical devices. In addition to influencing the mechanical modulus, the transition in nodal-topology can be harnessed to generate changes in other physical properties like thermal and electrical conductivity. This gives rise to various other applications such as sensors. 


\backmatter

\section*{Supplementary information}

Supplementary information is provided as a pdf.

\section*{Ethics declarations}

\subsection*{Competing interests}
The authors declare no competing interests.

\section*{Data availability}
Data supporting the findings of this study are available from the corresponding authors on request. 

\section*{Acknowledgements}
This study is part of the strategic focus area advanced manufacturing project Sustainable Design of 4D Printed Active Systems (SD4D).

\bibliography{References_top_mat}

\end{document}